\documentclass[prd,twocolumn,amsmath,amssymb,floatfix,superscriptaddress,nofootinbib,preprintnumbers]{revtex4-1}
\pdfoutput=1
\usepackage{color}
\usepackage{graphicx}
\usepackage{tabularx}
\usepackage{xspace}
\usepackage[justification=centerlast, format=plain, labelfont=bf]{caption}
\usepackage{subcaption}
\usepackage{float}
\usepackage{verbatim}
\usepackage{amsmath}
\usepackage{amssymb}
\usepackage{url}
\usepackage{bbold}
\usepackage{slashed}
\usepackage{array}
\usepackage{multirow}
\usepackage{threeparttable}
\usepackage{paralist}
\usepackage{xspace}
\usepackage{upgreek}
\usepackage{lipsum}
\usepackage[colorlinks]{hyperref}
\newcolumntype{C}{>{\centering\arraybackslash}X}
\newcommand{\taur}{{\tau_{\mathrm{reio}}}}

    \newcommand{\be}{\begin{equation}}
  \newcommand{\ee}{\end{equation}}
    \newcommand{\ba}{\begin{align}}
  \newcommand{\ea}{\end{align}}

\newcommand{\du}{{\rm d}}

\begin{document}

\title{Accurately Weighing Neutrinos with Cosmological Surveys}
\author{Weishuang Linda Xu}
\affiliation{Department of Physics, Harvard University, Cambridge, MA 02138, USA}
\author{Nicholas DePorzio}
\affiliation{Department of Physics, Harvard University, Cambridge, MA 02138, USA}
\author{Julian B.~Mu\~noz}
\affiliation{Department of Physics, Harvard University, Cambridge, MA 02138, USA}
\author{Cora Dvorkin}
\affiliation{Department of Physics, Harvard University, Cambridge, MA 02138, USA}

\begin{abstract}
A promising avenue to measure the total, and potentially individual, mass of neutrinos consists of leveraging cosmological datasets, such as the cosmic microwave background and surveys of the large-scale structure of the universe. In order to obtain unbiased estimates of the neutrino mass, however, many effects ought to be included. Here we forecast, via a Markov Chain Monte Carlo likelihood analysis, whether measurements by two galaxy surveys: DESI and {\it Euclid}, when added to the CMB-S4 experiment, are sensitive to two effects that can alter neutrino-mass measurements.
The first is the slight difference in the suppression of matter fluctuations that each neutrino-mass hierarchy generates, at fixed total mass.
The second is the growth-induced scale-dependent bias (GISDB) of haloes produced by massive neutrinos.  We find that near-future surveys can distinguish hierarchies with the same total mass only at the $1\sigma$ level; thus, while these are poised to deliver a measurement of the sum of neutrino masses, they cannot significantly discern the mass of each individual neutrino in the foreseeable future. 
We further find that neglecting the GISDB induces up to a $1\sigma$ overestimation of the total neutrino mass, and we show how to absorb this effect via a redshift-dependent parametrization of the scale-independent bias. 
\end{abstract}

\maketitle

\section{Introduction}

The existence of neutrinos has long been established, but comparatively little is known about them, due to their weak couplings to the visible sector. 
Although in the Standard Model (SM) neutrinos are massless, compelling evidence of flavor oscillations from solar, atmospheric, and reactor fronts~\cite{McDonald:2016ixn,Kajita:2016cak,Nunokawa:2007qh,GonzalezGarcia:2012sz,Fogli:2012ua,Maltoni:2002aw,Maltoni:2004ei} yield measurements of two mass splittings, indicating non-zero masses for at least two of the three neutrino species.  The sign of one of the measured mass splittings is yet to be determined, suggesting that neutrinos are ordered in one of two scenarios: the normal hierarchy (NH), where the two lighter neutrinos are closer in mass, or the inverted hierarchy (IH), where the two heavier ones are.  Distinguishing between the two neutrino hierarchies, as well as measuring their overall mass scale, are integral steps towards amending the Standard Model via characterizing its least-understood fermions.

Current results from the KATRIN Tritium decay experiment have improved the limits on the mass $m_\nu$ of each neutrino species to $m_\nu<1100$ meV in the quasi-degenerate regime~\cite{Aker:2019uuj}, and are expected to constrain each neutrino mass in this regime to $200$ meV with upcoming data~\cite{Osipowicz:2001sq}. 
This is, however, still far from the minimum (total) masses expected for the NH and IH, of 60 and 100 meV respectively. 
A diverse range of other particle experiments are also underway aiming to fully characterize the oscillation parameters and determine the mass ordering~\cite{An:2015jdp,Fukuda:1998rq,Collaboration:2011nsa,Abe:2011fz,Ayres:2004js,Ahn:2012nd,Abe:2011sj}. At the same time, cosmological data sets provide a powerful tool in the search for massive neutrinos, as these particles are very abundant in our universe, with a density per species today of $n_{\nu}\sim 100 \,\rm cm^{-3}$, comparable to that of cosmic microwave background (CMB) photons.  
This cosmic neutrino background influences the formation of large-scale structure (LSS) in the universe: at least two of these species are non-relativistic at the present day and contribute to the dark matter (DM) content. However, their small masses imply large streaming velocities and induce structure suppression at small scales (see Refs.~\cite{Lesgourgues:2006nd,Lattanzi:2017ubx} for detailed reviews of these effects). 

Indeed, the leading constraints on the sum of neutrino masses are currently obtained with cosmological data~\cite{Vagnozzi:2017ovm,Lattanzi:2017ubx,Castorina:2013wga, Allison:2015qca,Cuesta:2015iho}. 
The latest 2018 {\it Planck} data, in conjunction with measurements of the baryon acoustic oscillations (BAO) from the Baryon Oscillation Spectroscopic Survey (BOSS), have been used to constrain $\sum m_\nu \leq 120$ meV at 95\% C.L.~\cite{Aghanim:2018eyx}.  
This measurement is compatible with both the normal and inverted hierarchies, though the available parameter space for the latter will always be smaller. 
Upcoming data from the CMB Stage-4 (CMB-S4) experiment, as well as the Dark Energy Spectroscopic Instrument (DESI) and {\it Euclid} galaxy surveys, will reduce these error bars dramatically, and it is expected that these experiments will measure the sum of neutrino masses at least at the $3\sigma$ level~\cite{Dvorkin:2019jgs}.  

As cosmological data (especially that from large-scale structure experiments) become increasingly precise, it becomes critical to accurately characterize the cosmological effects of neutrinos in the analysis of this data.  
This is crucial both for the correct characterization of cosmological neutrinos and also for the measurement of relevant parameters of structure formation, such as the amplitude of fluctuations or the intrinsic bias of tracers. Currently, searches for massive neutrinos with cosmological data often make two simplifying assumptions, described below.

The first assumption commonly made is that the three neutrinos have the same mass, a configuration commonly termed the degenerate hierarchy (DH). While this is a valid approximation for neutrino masses much heavier than their splittings ($m_{\nu_i}\gg 50$ meV), it fails for the range of masses still allowed by current data. While the dominant cosmological effect of neutrinos is set by the sum of their masses, the distribution of individual masses has an effect that, although subtle, might be detectable by future surveys~\cite{Lesgourgues:2004ps,Pritchard:2008wy,DeBernardis:2009di,Jimenez:2010ev,Hall:2012kg,Giusarma:2013pmn,Hannestad:2016fog,Vagnozzi:2017ovm,Archidiacono:2020dvx}. It is possible that the next-generation measurement of total neutrino mass will simply eliminate the inverted hierarchy by ruling out its minimum mass. However, in the case that this measured sum permits both normal and inverted configurations  ($\sum m_\nu \geq 100$ meV), it is also worth investigating whether cosmological data has the power to distinguish between the two.

The second approximation is that the halo bias is unaltered by the presence of light degrees of freedom. 
However, it was shown in Refs.~\cite{LoVerde:2014rxa,LoVerde:2014pxa,Munoz:2018ajr} that the same scale-dependent growth that gives rise to a suppression in the matter power spectrum in the presence of massive neutrinos induces a scale- and redshift-dependent enhancement to the halo bias as well. 
This growth-induced scale-dependent bias (GISDB) can partially compensate the effect of neutrinos on the galaxy power spectrum and, opposed to other biasing effects that abound in standard cosmology, its amplitude and shape are determined by the neutrino masses, so it must be properly modeled during searches for these particles. 

In this work we include both of these effects for the first time (see for instance Refs.~\cite{Raccanelli:2017kht, Vagnozzi:2018pwo, LoVerde:2016ahu} for previous efforts), and forecast constraints on the neutrino mass from the upcoming DESI~\cite{Aghamousa:2016zmz} and {\it Euclid}~\cite{Amendola:2016saw} surveys, combined with the CMB-S4 experiment~\cite{Abazajian:2016yjj}. 
We study whether the omission of these corrections would bias upcoming results.
To find the halo power spectrum in the presence of neutrinos with three different masses we employ the publicly available software {\tt CLASS}~\cite{Blas:2011rf}, which we have modified to include the recently developed {\tt RelicFast}\footnote{\url{https://github.com/JulianBMunoz/RelicFast}}~\cite{Munoz:2018ajr} code as a native module. We dub this code {\tt RelicCLASS}\footnote{\url{https://github.com/wlxu/RelicClass}}, and in addition to neutrino masses, it can also be used to search for any other light relic (as we do in our companion paper \cite{DePorzio:2020wcz}).

This paper is structured as follows: in \S~\ref{sec:effects} we briefly review the effect of neutrinos on the LSS observables. In \S~\ref{sec:methods} we explain the datasets we consider, which we employ in \S~\ref{sec:results} to forecast the resulting constraints on neutrino masses. We conclude in \S~\ref{sec:conclusions}.

\section{Neutrinos and their effect on the LSS}
\label{sec:effects}

We begin with an overview of the physics of neutrinos and their effects on the LSS observables, which can be divided in two pieces: the suppression of the matter power spectrum, and a modification to the galaxy-halo bias. Both of these effects are most relevant at scales of $k\sim 0.01$ $h/{\rm Mpc}$ for currently allowed neutrino masses, making galaxy surveys such as DESI and {\it Euclid} ideal probes, given their expected low noise at those scales.

The SM contains three species of neutrinos corresponding to the electron, muon, and tau lepton flavors, which mix into three non-degenerate mass eigenstates. 
Various oscillation experiments have measured two mass splittings to be~\cite{Maltoni:2004ei}
\begin{align}
\Delta m^2_{21} &= 79\, \mathrm{meV}^2 \nonumber\\
|\Delta m^2_{31}| &= 2.2 \times 10^3 \, \mathrm{meV}^2,          
\end{align}
where the absolute value on the latter measurement denotes ambiguity to which species is heavier. This results in two possible mass configurations, the normal hierarchy, where $m_3 > m_1$, and the inverted hierarchy, where $m_3 < m_1$. Assuming the lightest neutrino is massless, the NH has a total mass sum of $\sim 60$ meV and the IH has one of $\sim 100$ meV~\cite{Lesgourgues:2006nd}. 

Neutrinos decouple shortly before cosmic electron-positron annihilation, and so their present-day temperature $T_\nu^{(0)}$ is offset from that of the photon bath, $T^{(0)}_\gamma$, by the subsequent entropy injection, yielding  $T_\nu^{(0)}=(4/11)^{1/3} T^{(0)}_\gamma = 1.95$ K. 
Thus, a neutrino of mass $m_{\nu_i}$ will become nonrelativistic at $z_{\rm NR} \sim 500  (m_{\nu_i}/100 \mathrm{meV})$, so by today we expect at least two of the species to be non-relativistic.
In that case, the neutrino abundance $\Omega_\nu$ today is related to the sum of masses simply as~\cite{Lesgourgues:2006nd}
\be
\Omega_\nu h^2= \sum_i  \dfrac{m_{\nu_i}}{93.2\, \rm eV},
\ee
where $h$ is the reduced Hubble parameter.
The non-zero temperature of neutrinos allows them to freely stream out of dark-matter structure. This defines a free-streaming scale as the wavenumber $k_{\mathrm{fs}, i}$ above which neutrinos behave as hot dark matter, given at late times $z< z_{\mathrm{NR},i}$  by
\be
k_{\mathrm{fs}, i} = \frac{0.08}{\sqrt{1+ z}} \left( \dfrac{m_{\nu_i}}{100 \rm meV}\right)  h\,\rm Mpc^{-1},
\ee
assuming matter domination~\cite{Bird:2018all,AliHaimoud:2012vj}.
We will focus on neutrinos in this paper, but we note that other light (but not massless) relics produce similar effects, and we search for them in our companion paper \cite{DePorzio:2020wcz}.

\subsection{Effect on Matter Fluctuations}

For this discussion it is informative to differentiate between two different types of matter content in the universe: the component consisting of cold dark matter and baryons (CDM+b), which largely follow each other at late times and actually source the formation of galaxies, and neutrinos, which are generally non-relativistic at the present day but do not cluster at small scales~\cite{Biagetti:2014pha,Castorina:2015bma}. 
At late times, when all neutrinos are non-relativistic, we can define the matter fluctuations as a sum of these two components:
\be
\delta_m = f_{\rm cb} \delta_{\rm cb} + \sum_i f_{\nu_i} \delta_{\nu_i},
\label{eq:deltam}
\ee
where $\delta_i$ is the overdensity in the $i$ component, $f_i=\Omega_{i}/\Omega_m$ its fraction of the total matter abundance $\Omega_m$, and we define $f_\nu = \sum_i f_{\nu_i}$, where this sum includes all massive neutrinos.

On large scales ($k \ll k_{\mathrm{fs},i}$), neutrinos $\nu_i$ will follow CDM+b fluctuations, so $\delta_\nu \approx \delta_{cb}$, whereas on small scales ($k \gg k_{\mathrm{fs},i}$) they will freely stream out of matter potential wells and their fluctuations will be suppressed, following $\delta_\nu \propto \delta_{cb} k^{-2}$~\cite{Lesgourgues:2006nd}. 
This affects structure formation in two main ways. 
First, $\delta_m$ is suppressed by a factor of $(1-f_{\nu})$ with respect to $\delta_{cb}$, as the larger the neutrino abundance, the smaller the fraction of matter content that contributes to the growth of structure. 
Second, and more important, the absence of small-scale neutrino fluctuations slows the growth of the CDM+b component at large $k$. 
If the neutrinos become non-relativistic after matter-radiation equality, this produces an additional scale-dependent suppression on $\delta_{cb}$ of roughly $(1- 3f_{\nu})$. The result in the linear approximation is a total suppression of the matter power spectrum of $(1- 8 f_\nu)$ for $f_\nu \ll 1$~\cite{Lesgourgues:2006nd}. 
 
Both of these effects become present at $k\sim k_{\mathrm{fs},i}$ for each species, so not only do neutrino masses determine the overall amount of suppression, but also the location in the power spectrum where said suppression occurs. The effect is dominantly determined by the total $f_\nu$, and thus the total neutrino mass $\sum m_\nu$. However, hierarchical neutrinos will each modify the power spectrum at slightly different free-streaming scales due to their individual masses. For instance, for the same total neutrino mass of $\sum m_\nu = 100$ meV, the normal hierarchy suppression effects turn on at $k_{\mathrm{fs},i} \sim 0.02 \, h$ ${\rm Mpc}^{-1}$ (for the lighter neutrinos) and $k_{\mathrm{fs},i} \sim 0.04 \,h\,{\rm Mpc}^{-1}$ (for the heavier), while the inverted hierarchy suppression effects turn on at $k_{\mathrm{fs},i} \sim 0.005 h\,{\rm Mpc}^{-1}$ and $k_{\mathrm{fs},i} \sim 0.03h {\rm Mpc}^{-1}$,  respectively. 

Furthermore, although the difference is subtle, the amplitude of the small-scale suppression of $P_{cb}(k)$ for cosmologies with fixed total  $f_\nu$ is dependent on the epochs $z_{\mathrm{NR}, i}$ where neutrinos begin to behave as matter, with larger suppression for heavier individual species as the growth of fluctuations is slowed from an earlier time. Thus, the scale-dependent suppression is most prominent in the limit where all the mass is carried by one neutrino, and least prominent for the case of three degenerate neutrino; and in general, the inverted scenario will generate more suppression for the same total neutrino mass than the normal one.

In addition to these scale-dependent effects, the inclusion of massive neutrinos while holding fixed the baryon and DM abundances ($\omega_b,\, \omega_{\rm cdm}$), and the Hubble parameter ($h$) forces a shift in the dark-energy abundance ($\Omega_\Lambda$), the effect of which is an overall suppression of the amplitude of fluctuations at all scales. However, this effect is less important for our analysis here, as it can be mimicked by a compensating shift in the amplitude $A_s$ of fluctuations, the Hubble parameter, or the overall halo bias.

\begin{figure}

\includegraphics[width = \linewidth]{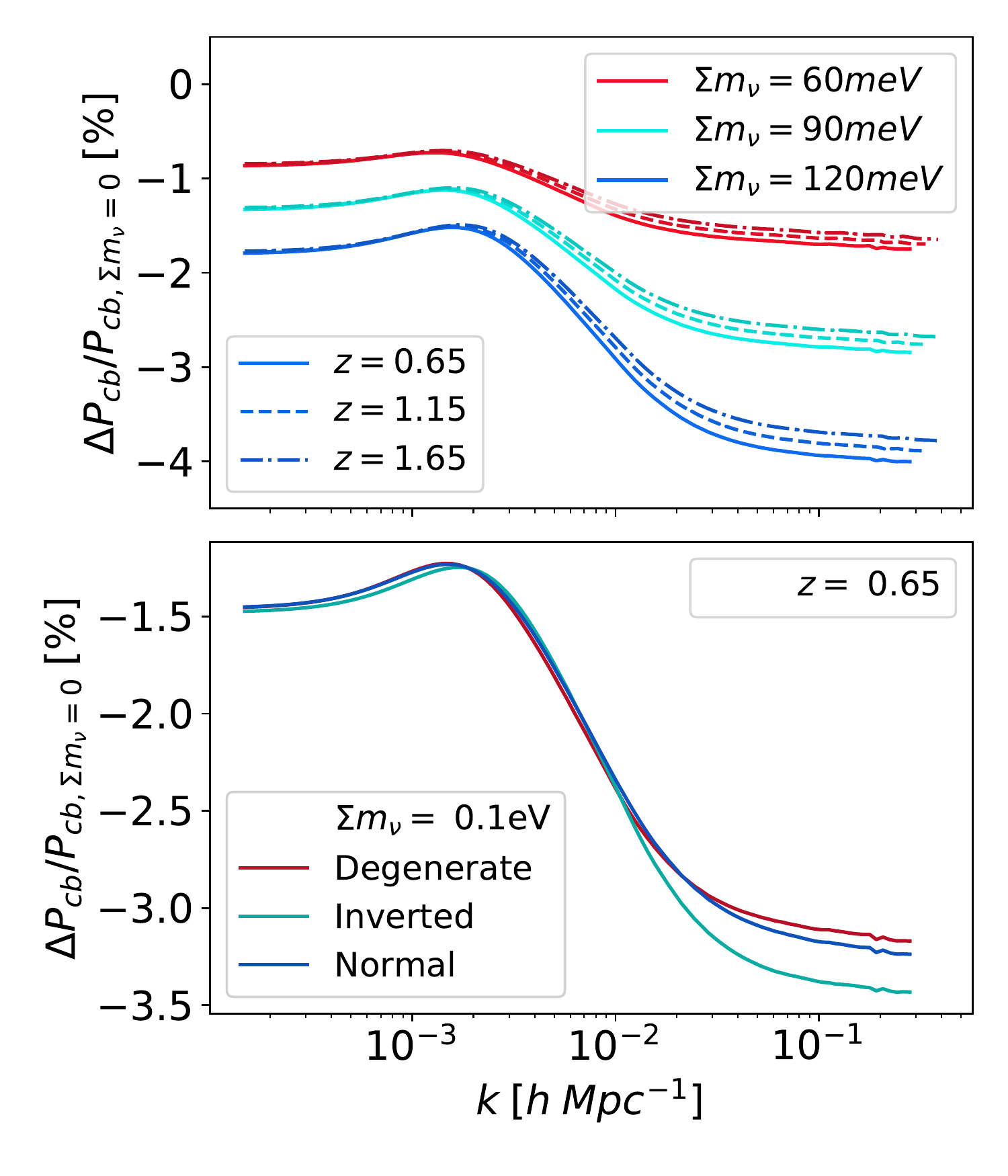}
\caption{Percent differences in CDM+baryon power spectra compared to a massless neutrino cosmology; for different total neutrino masses assuming the degenerate hierarchy (upper panel) and for different hierarchies assuming a total neutrino mass of 100 meV (lower panel). In each case we fix the cosmological parameters $\{\omega_b, \omega_{\rm cdm}, \Omega_m, A_s, n_s\}$, varying $h$. 
As shown, the primary effect of massive neutrinos is a suppression of amplitude at small scales -- the change in amplitude at large scales is attributed to varying values of $h$. Note that both the total mass and individual neutrino masses affect the amount of suppression and scale at which it turns on, though the latter effect is subdominant. In addition, the amount of suppression is redshift dependent, with a larger spread at small scales for larger neutrino masses.
} 
\label{fig:suppression_all}
\end{figure}

We illustrate the effects described above in Figure~\ref{fig:suppression_all}, where we incorporate massive neutrinos into the cosmology with various masses and fixed hierarchy (upper panel) and with various hierarchies and fixed total mass (lower panel). We investigate the suppression of each with respect to a cosmology with massless neutrinos, holding fixed the other cosmological parameters $\{ \omega_b, \omega_{\rm cdm}, h, A_s, n_s, \taur \}$ at values listed in Table~\ref{tab:fiducials} (where $n_s$ is the tilt of primordial fluctuations and $\taur$ is the optical depth to reionization).  Note that the suppression is larger for heavier neutrinos and occurs at smaller scales, and also it is larger for the inverted hierarchy than the normal hierarchy, though only becoming apparent at comparatively small scales.

\subsection{Effect on the Bias -- {\tt RelicCLASS}}

We can observe the neutrino-induced suppression in $\delta_m$ by directly measuring the matter power spectrum, for example through weak lensing of the CMB or galaxies~\cite{Kaplinghat:2003bh,Abbott:2017wau}.
Most measurements are, however, of galaxy distributions, which trace the underlying matter fluctuations.
In this case it is not enough to study how neutrinos affect the matter power spectrum, but rather it is necessary to find how they change the relation between the halo and the matter overdensities, the galaxy bias.

Due to the nonlinearity of halo formation, the overdensity of haloes traces that of the matter, albeit with a rescaling referred to as the bias. In this work we will always refer to the bias with respect to the CDM+b field, to avoid spurious scale-dependencies due to the non-clustering nature of neutrinos~\cite{Biagetti:2014pha,Villaescusa-Navarro:2013pva,Castorina:2013wga,Costanzi:2013bha}. 
In that case, the halo fluctuation (without redshift-space distortions) is given by
\be
\delta_h(k,z) = b_1(k,z) \delta_{cb}(k,z)
\ee
to linear order, where $b_1$ is the Eulerian bias, which can be written in terms of the linear Lagrangian bias $b^L$ as
\be
b_1(k,z) = 1 + b^L(k,z),
\ee
where the Lagrangian bias is also defined with respect to the $cb$ fluid. 

In previous LSS searches for neutrino masses it was typically assumed that either $b_1$, or its equivalent with respect to all matter, was constant at all scales. 
Nonetheless,  neutrinos produce a scale- and redshift-dependent growth in the CDM+b fluctuations, due to their free-streaming nature.  This effect cannot be simply included through transfer functions, due to the non-local temporal nature of the halo-formation process~\cite{Senatore:2017hyk,Senatore:2014eva,LoVerde:2014pxa,LoVerde:2014rxa}, and the process of halo collapse has to be modeled. 

We use the publicly available code {\tt RelicFast}, which solves for the spherical collapse of haloes including the effect of neutrinos. In Ref.~\cite{Munoz:2018ajr} we found that, while the overall value of the bias is very sensitive to the astrophysics of specific tracers, the scale-dependence of the Lagrangian bias is impervious to those effects within our model.

\begin{figure}
    \includegraphics[width = \linewidth]{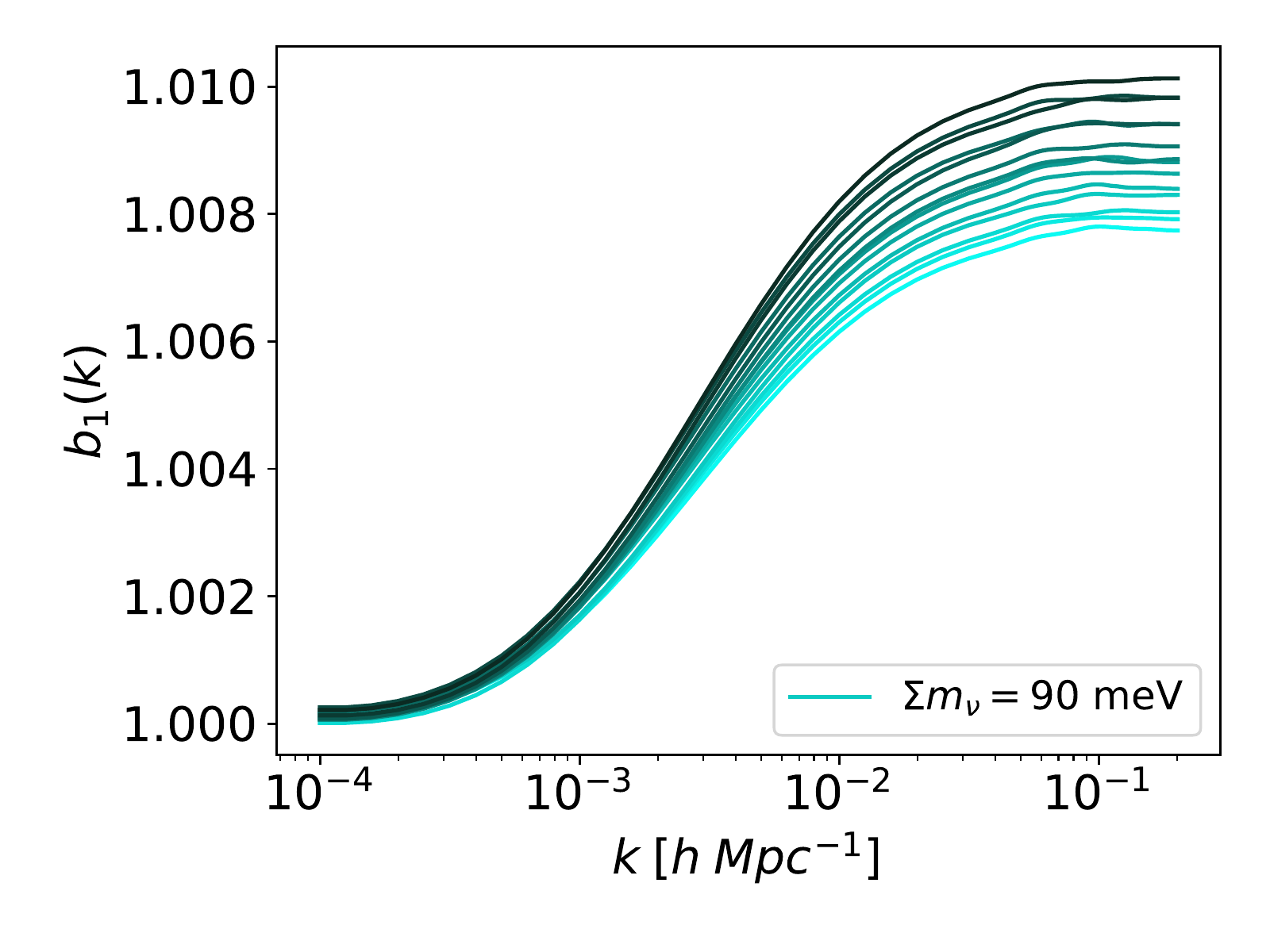}
    \caption{The growth-induced scale-dependent bias (GISDB) for redshifts from 0.65 (lightest) to 1.65 (darkest) with massive neutrinos. The total neutrino mass is set at 90 meV and the degenerate scenario is assumed. As shown, the growth-induced scale-dependent bias (GISDB) is both scale- and redshift-dependent. 
    }
    \label{fig:bias}
\end{figure}

The result of this correction is a scale-dependent step in the Eulerian bias as shown in Figure~\ref{fig:bias}, which we term the \textit{growth-induced scale-dependent bias} (GISDB).
We parametrize the bias through
\be
b^L(k,z) = \overline {b^L}(z) f(k,z),
\ee
where $f(k,z)$ is numerically computed with {\tt RelicFast} to properly account for the effect of neutrinos in the halo bias, and $\overline {b^L}(z)$ denotes the scale-independent magnitude of the Lagrangian bias, which we will marginalize over. 
We remind the reader that $f(k)$ can be approximated in terms of $\tanh[\log(k)]$, as it is roughly a step function in log-$k$ space~\cite{Munoz:2018ajr}. Nonetheless, we choose to use the full shape of the function, in order to fully capture its physical effect. 
Additionally, as the amplitude $\overline {b^L}(z)$ of the bias depends very sensitively on the properties of the haloes studied, we will marginalize over it as a free parameter.
On the other hand, the scale-dependent behavior, parametrized through $f(k,z)$, is largely independent of those factors~\cite{Munoz:2018ajr,LoVerde:2014pxa}.

The specifications of the {\tt RelicCLASS} code is as follows: We have modified the publicly available Boltzmann solver {\tt CLASS} to include the effect of neutrinos in the halo bias computed by {\tt RelicFast} within {\tt CLASS}, which directly outputs both Eulerian and Lagrangian scale-dependent biases for the input cosmologies. Here, {\tt RelicFast} is included as a {\tt CLASS} module, executed after the {\tt Lensing} module, which outputs the realistic scale-dependent Lagrangian bias for requested ranges of redshifts and halo masses. This output is accessible from the python wrapper to facilitate interfacing with {\tt MontePython} and other codes that take {\tt CLASS} outputs. Inverted, normal and degenerate neutrino scenarios with lightest neutrino mass ($m_0$ in the code), which sets the full spectrum, are accepted as specifications at the input level. Aside from neutrinos, {\tt RelicCLASS} can also be used to model other scale-dependent effects on the growth function, such as those induced by other light (but massive) degrees of freedom.

\section{Datasets}
\label{sec:methods}

In this section we describe the data sets used in our analysis, code specifications, and discuss details of likelihoods and nuisance parametrizations. We use mock data from CMB-S4 as well as either DESI or {\it Euclid} for the LSS component. 

\subsection{Galaxy Data}

For the LSS component we will use a modified version of the basic {\tt pk} likelihood implemented in {\tt Montepython v3}~\cite{Brinckmann:2018cvx} adapted to mock data from the upcoming DESI~\cite{Aghamousa:2016zmz} and {\it Euclid}~\cite{Amendola:2016saw} surveys, assuming their most abundant tracers. For this analysis we consider emission line galaxies (ELGs) for DESI, and assume the baseline survey covering 14,000 $\rm{deg}^2$. 
Conversely, for {\it Euclid} we study H$\alpha$ emitters and assume the reference efficiency given in Ref.~\cite{Amendola:2016saw} with 15,000 $\rm{deg}^2$ coverage. 
The projected galaxy number densities achievable by the DESI and {\it Euclid} surveys are given in Table~\ref{tab:ELGs}. Somewhat more optimistic constraints could potentially be achieved through multi-tracer techniques~\cite{Boyle:2017lzt}.
Nonetheless, our goal in this work is to determine whether the inclusion of different neutrino-induced effects would bias the results from upcoming surveys, so we will limit ourselves to the case of one tracer per survey.

\begin{table*}[t!]
    \begin{center}
    \vspace{3mm}
        \begin{tabularx}{\textwidth}{ c | CCCCCCC  }
        \toprule
       $z$  & 0.65  & 0.75 & 0.85 & 0.95 & 1.05 & 1.15  & 1.25  \\ 
       \hline \hline 
      \rule{0pt}{4ex}
      $\displaystyle{\frac{\du N_{ELG}}{\du z \; \du\mathrm{deg}^2}}$ [DESI] & 309 & 2269 & 1923 & 2094 & 1441 & 1353 & 1337  \\
      \cline{2-8}
       \rule{0pt}{4ex} $\displaystyle{\frac{dN_{H\alpha}}{\du z \; \du\mathrm{deg}^2}}$ [{\it Euclid}] & 2434 & 4365 & 4729 & 4826 & 4729 & 4508 & 4270 \\ 
      \botrule
    \end{tabularx}
    
    \vspace{3mm}
    \begin{tabularx}{\textwidth}{c  |  CCCCCCC }
        \toprule
       $z$  &  1.35 & 1.45 & 1.55 & 1.65  &  1.75 & 1.85 & 1.95 \\ 
       \hline \hline
     \rule{0pt}{4ex}
      $\displaystyle{\frac{\du N_{ELG}}{\du z \; \du\mathrm{deg}^2}}$ [DESI]&  523 & 466 & 329 & 126 & 0 & 0 & 0 \\
      \cline{2-8}
       \rule{0pt}{4ex} $\displaystyle{\frac{dN_{H\alpha}}{\du z \; \du\mathrm{deg}^2}}$ [{\it Euclid}] &  3721 & 3104 & 2309 & 1514 & 1475 & 894 & 498\\ 
      \botrule
    \end{tabularx}
\end{center}
    \caption{Forecasted number of ELGs measurable by DESI  and H$\alpha$ emitters measurable by {\it Euclid} per redshift per $\rm{deg}^2$ at each redshift bin $z$, taken from Refs.~\cite{Aghamousa:2016zmz,Amendola:2016saw}.}
    \label{tab:ELGs}
\end{table*}

The shot noise spectrum is given simply by the inverse of the observed galaxy density,
\be
\overline n_g^{-1}(z)  = \dfrac{dV(z)}{dN_g(z)},
\ee
where $dV(z)$ is the comoving volume of the shell at redshift $z$ in the fiducial cosmology, and $dN_g(z)$ is the total number of tracer galaxies within the shell, as computed in Table~\ref{tab:ELGs}.

Galaxies are located in the line-of-sight $\hat n$ direction at their measured redshift $z$.  Gravitational attraction of galaxies into clusters, as well as bulk velocities of the clusters themselves, give rise to distortions to the inferred 3D positions of galaxies, which are usually termed redshift-space distortions (RSD)~\cite{Kaiser:1987qv}.
To linear order we can relate the redshift-space galaxy power spectrum to the CDM+b one as
\be
P_g(k,z,\mu) = \left[b_1(k,z) + f_{cb}(k) \mu^2\right]^2 P_{cb}(k,z),
\label{eq:PgRSD}
\ee
where $\mu=\hat k \cdot \hat n$, and we have defined the  growth factor of CDM+b fluctuations as
\be
f_{cb}(k,z) = -\dfrac{d \log \sqrt{P_{cb}(k,z)}}{d \log z},
\ee
which we compute numerically using {\tt CLASS}.

Additionally, the non-linear integrated effect of RSD (usually referred to as the Finger-of-God effect), as well as the intrinsic redshift uncertainty of the galaxy, can be encoded as a multiplicative damping term~\cite{Bull:2014rha}
\begin{align}
\tilde P_g (k,z,\mu) = P_g (k,z,\mu) \exp\left[-\frac{\mu^2 k^2 \sigma_v^2(z)}{H^2(z)}\right], \nonumber \\{\rm with} \quad
\sigma_v = (1+z)\sqrt{\sigma_0^2 +  \frac{\sigma^2_{\rm fog} (1+z)}{2}}, 
\label{eq:tildePg}\end{align}
where $\sigma_0 = 10^{-3}$ accounts for the resolution limits of DESI~\cite{Aghamousa:2016zmz} and {\it Euclid} \cite{Amendola:2016saw}, and $\sigma_{\rm fog}$ is related to the velocities of galaxies and is treated as a nuisance parameter in this work. 

The physics of galaxy formation is known to produce an additional scale-dependence to the bias term, proportional to $k^2$ at high $k$. While unrelated to the effect of neutrinos on the LSS, this $k^2$-term is properly accounted for here following the formulation of Ref.~\cite{Modi:2016dah}. We follow their prescription in writing the total Eulerian bias as  
\be
b_1(k,z) = 1 + b^L(k,z) + \alpha_2 k^2,
\label{eq:biasE}
\ee
where $\alpha_2$ is a free parameter that we marginalize over. At large scales $(k\to0)$, where both the growth-induced and $k^2$-terms are negligible, the bias is scale-independent and we choose a fiducial value that matches the simulations of each specific tracer. 
For the DESI survey of ELGs we use the parametrization $b_1(z) = \beta_0/D(z)$, where $D(z)$ is the growth function, and for the {\it Euclid} survey of H$\alpha$ emitters we use $b_1 (z)  = \beta_0 (1+z)^{0.5\beta_1}$, following the prescriptions of their respective Science Books~\cite{Aghamousa:2016zmz,Amendola:2016saw}.  The nuisance parameter $\beta_0$ rescales the overall bias, and $\beta_1$ parametrizes any uncertainty in redshift dependence of the bias. We will take as fiducial $\beta_{0, \rm DESI} =1.0,\; \beta_{0,   Euclid} =1.7,\;  \beta_1 = 1.0$, consistent with recent results from simulations \cite{Chiang:2018laa}. 

\begin{figure}
    \includegraphics[width = \linewidth]{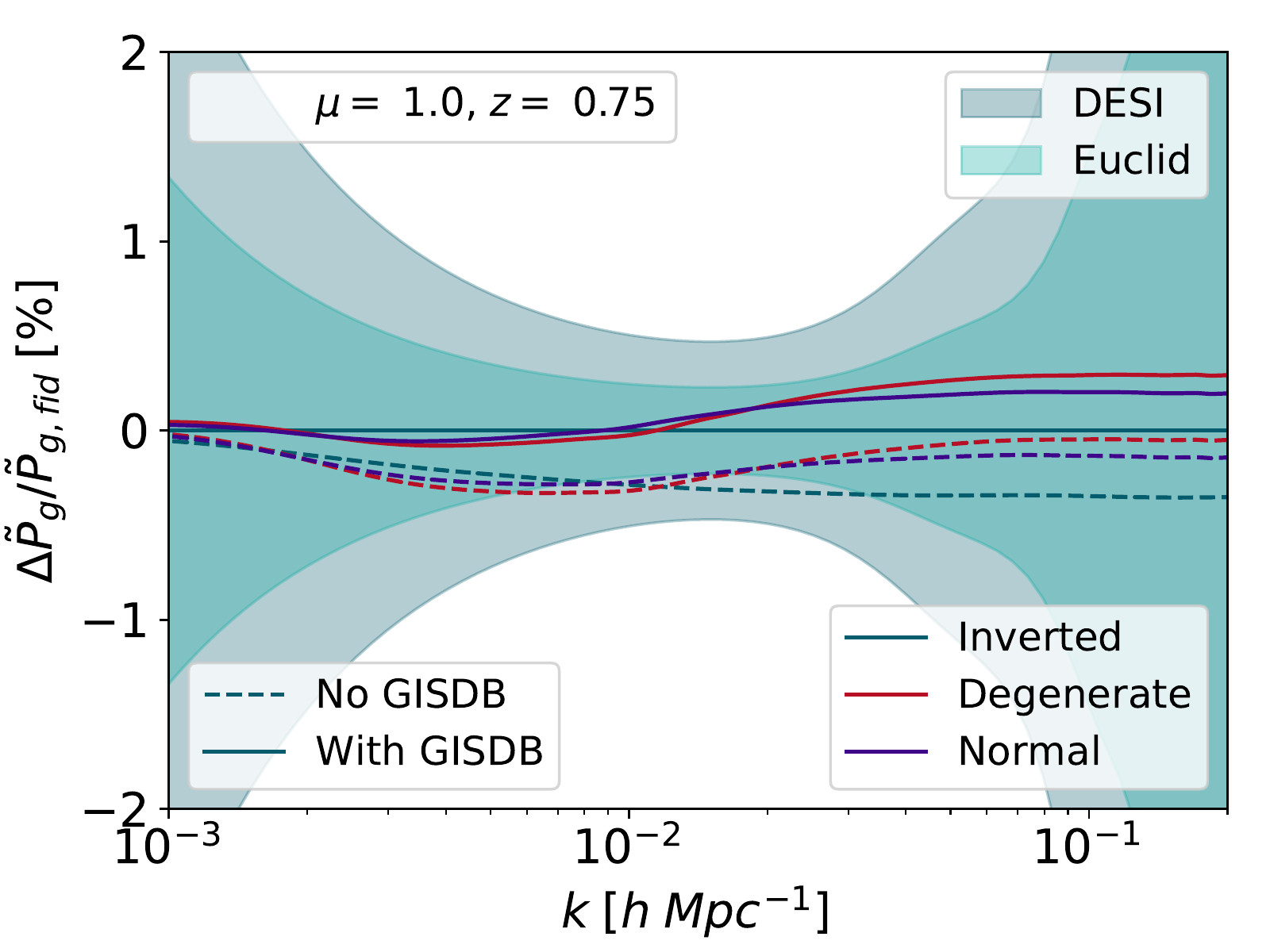}
    \caption{Percent differences in galaxy power spectra $\tilde P_g(k,z,\mu)$ between the various neutrino hierarchies (at fixed $\sum m_\nu = 100$ meV), as well as with and without the GISDB, compared to a fiducial case of inverted hierarchy with GISDB, at $z=0.75$. The shot noises associated with DESI and {\it Euclid} are shown as the shaded areas. Here the cosmological parameters $\{\omega_b, \omega_{\rm cdm}, h, A_s, n_s,  \sum m_\nu\}$ as well as all bias and RSD nuisance parameters are held fixed.}
    \label{fig:desi_pk}
\end{figure}

Figure~\ref{fig:desi_pk} shows the percent differences in $\tilde P_g(k,z,\mu)$ with respect to a fiducial scenario of inverted neutrino hierarchy with $\sum m_\nu = 100$ meV, upon changing the hierarchy (while fixing the total neutrino mass) and switching off the GISDB. Other cosmological parameters are held fixed, and the shaded regions represent the shot noises expected from DESI and {\it Euclid}. 
This figure shows that the effect of the neutrino hierarchies, as well as the GISDB, is at the 0.5\% level.
However, while the hierarchies affect the ratio differently at different scales, the GISDB acts as an overall change in normalization at $k\gtrsim 10^{-3}\,h/\rm Mpc$.

We also account for the Alcock-Paczynski effect~\cite{Alcock:1979mp,Asgari:2016txw,Lemos:2017arq}, which concerns the cosmology-dependence of inferring distance from angular- and redshift-space measurements, by multiplying the power spectrum and the shot noise at each bin by $H(z)/D_A^2(z)$, dividing by the same quantity evaluated at our fiducial cosmology, as well as writing the inferred $k$, $\mu$ with respect to the fiducial $k$, $\mu$ by the relation

\be \frac{k(z, \mu_{\rm{fid}})}{k_{\rm{fid}}} =  \left[(1-\mu_{\rm{fid}}^2) \frac{D_{A, \rm{fid}}^2(z)}{D_{A}^2(z)} + \mu_{\rm{fid}}^2 \frac{H^2(z)}{H_{\rm{fid}}^2(z)}\right]^{1/2}, \ee

\be \frac{\mu(z)}{\mu_{\rm{fid}}} =   \left[(1-\mu_{\rm{fid}}^2) \frac{D_{A,\rm{fid}}^2(z)}{D_{A}^2(z)} \frac{H^2_{\rm{fid}}(z)}{H^2(z)}  + \mu_{\rm{fid}}^2 \right]^{-1/2}. \ee
Additionally, we stop our analysis at $k_{\rm max}(z) =0.2\times (1+z)^{2/(2+n_s)} h $ Mpc$^{-1}$, safely within the linear regime. 

We compute the log-likelihood as~\cite{Sprenger:2018tdb}

\begin{align}
- \log \mathcal L &= \frac{1}{2} \sum_z \int \du k_{\rm fid}  k_{\rm fid}^2 \int \du \mu_{\rm fid} \frac{V_{\rm fid}(z)}{2 (2\pi)^2}  \nonumber\\
&\left[\frac{\frac{H(z)}{D^2_A(z)}\tilde P_g(k,z, \mu)- \frac{H_{\rm fid}(z)}{D^2_{A, \rm fid} (z)}\tilde P_{g,\rm fid}(k_{\rm fid},z, \mu_{\rm fid})}{\frac{H(z)}{D^2_A(z)}\tilde P_g(k,z, \mu) + \frac{H_{\rm fid}(z)}{D^2_{A, \rm fid}} n_g^{-1}(z) }\right]^2.
\end{align}

Additional parameters, such as those accounting for non-Poissonian shot noise or theoretical error in this likelihood function, can be considered for more detailed analyses, as in e.g. Ref.~\cite{Brinckmann:2018owf}.

\subsection{CMB Data}

We complement the LSS information from galaxy surveys with mock CMB data from the upcoming CMB-S4 experiment, implemented with {\tt MontePython}'s {\tt Likelihood\_mock\_cmb}.
We model the CMB-S4 simply as a single effective frequency channel, with temperature noise $\Delta_T=1\,\mu$K-arcmin, and polarization noise $\Delta_P=\sqrt{2}\Delta_T$.
We additionally assume a resolution of $\theta_{\rm FWHM}=3$ arcmin.

CMB data will not only help break the degeneracies between cosmological parameters, but can also measure the matter power  spectrum directly through CMB lensing. 
This data will, for instance, break the degeneracy between the Hubble parameter and $\sum m_\nu$. We perform iterative delensing as in Ref.~\cite{Hirata:2003ka,Okamoto:2003zw} to obtain the deflection field with nearly optimal noise. Finally, we do not account for modes  below $\ell=30$ from CMB-S4, as it will be ground-based, and instead add
a Gaussian prior on $\taur$ with a width of 0.01 to account for low-$\ell$ CMB data. This width is reflective of current {\it Planck} sensitivities~\cite{Aghanim:2018eyx} but conservative in light of future measurements.

\section{Results}
\label{sec:results}

In this section we perform Markov Chain Monte Carlo (MCMC) analyses on mock data for CMB-S4 added to either DESI or {\it Euclid} galaxy power spectra, for cosmologies with massive neutrinos. We consider parameters $\{\omega_b, \omega_{\rm cdm}, h, A_s, n_s, \tau_{\rm reio}, \sum m_{\nu}\}$ as well as nuisance parameters $\{ \beta_0, \beta_1, \alpha_{k2}, \sigma_{\rm fog} \}$, and show our fiducial values in Table~\ref{tab:fiducials}.  
We vary our model in three types of ways.
First, we attempt to distinguish between the normal, inverted, and degenerate hierarchies.
Second, we study if the recovered parameters would be shifted if the GISDB was not included, both with and without marginalizing over the redshift-dependence uncertainty $\beta_1$. 
Additionally, we omit the $\taur$ prior in one case with CMB-S4+{\it Euclid}, to investigate the importance of additional optical depth information in the presence of the existing CMB lensing and LSS shape information. 

These different runs are designed to explore the set of physical effects that will appear in upcoming measurements of neutrino masses.
A table of relevant reconstructed parameters and associated best-fit log-likelihoods for selected models is shown in Table~\ref{tab:postparams_eu} for runs with {\it Euclid} data, and in Table~\ref{tab:postparams_desi} for those with DESI. We emphasize that in all cases our fiducial model corresponds to the inverted hierarchy with its lightest neutrino taken as massless, and thus represents a plausible model of nature; shifts from the best-fit in other models can be seen as the expected shift one would observe in a realistic analysis.

We find that with CMB-S4+DESI data the total neutrino mass is expected to be measured up to uncertainty of $26$ meV, while for CMB-S4+{\it Euclid} data that would be improved to $20$ meV. 
For $\sum m_\nu = 98$ meV this corresponds to a $4\sigma$ and $5\sigma$ detection respectively, and the \textit{minimum-mass} scenarios of normal and inverted hierarchy (with total masses $60$ and $100$ meV) can be distinguished at the $1.5\sigma$ and $2\sigma$ level. We note that neglecting the nuisance parameter $\beta_1$ results in an over-tightening  of $\sum m_\nu$ resolution to an uncertainty of 15 meV. Finally, the omission of a $\taur$ prior results in a $\sum m_\nu$ uncertainty of 27 meV for the combined CMB-S4+{\it Euclid} data; equivalently showing that these data are able to measure $\taur$ to the $10\%$ level despite the lack of low-$\ell$ information. Conversely, we find that a stricter $\taur$ prior of width 0.006 would tighten e.g. CMB-S4+DESI sensitivities to $20$ meV. 

\begin{table*}[]
    \begin{center}
    \begin{tabularx}{\textwidth}{ c | C |  c |  c |  c | c | C | C   | c |  c | c | c | C }
    \toprule
     \multicolumn{8}{c| }{Cosmological Parameters} & \multicolumn{5}{c}{Nuisance Parameters} \\ 
     \hline 
          \multicolumn{8}{c| }{}  & \multicolumn{2}{c |}{{\it Euclid}} &  \multicolumn{1}{c|}{DESI}  & \multicolumn{2}{c }{}  \\
         $\omega_b$  & $\omega_{\rm cdm}$ & $h$ & $A_s$ & $n_s$ & $\taur$ &  $\sum m_\nu$ [meV] & Hierarchy &   $\beta_0$ & $\beta_1$ & $ \beta_0 $ &  $\alpha_2$ & $\sigma_{\rm fog}$ [km/s] \\
         \hline 
         2.226e-2& 0.1127  & 0.701 &  2.2321e-9 &  0.967 & 0.0598 & 98.5 & Inverted &  1.7 & 1.0 & 1.0 & 1.0 & 250 \\
         \botrule
    \end{tabularx}
    \caption{Fiducial cosmology used in generation of mock data for MCMC analysis. Consistent cosmologies are used for DESI and {\it Euclid} analyses except for $\beta_0$, which is matched to simulation results.}
    \label{tab:fiducials}
    \end{center}
\end{table*}

\begin{table*}[]
\begin{tabularx}{\textwidth}{ c | c | c |  c | C | C | c | c | C | C }
    \toprule
     \multicolumn{2}{c|}{Data} & \multicolumn{3}{c|}{Model} & 
     \multicolumn{5}{c}{Mean and error} \\
     \hline 
     LSS & CMB & Hierarchy & Nuisance & GISDB  & $-2 \Delta \log \mathcal{L}$ & $\sum m_\nu$ [meV] & $\taur$ & $\beta_0 $ & $\beta_1$   \\
     \hline 
   \multirow{11}{*}{{\it Euclid}} & \multirow{10}{*}{CMB-S4  + $\taur$} & Degenerate  &$\{ \beta_0, \beta_1\}$ & Yes &  1.3  & 103.6 $\pm$  &  5.85e-2 $\pm$   &  1.702 $\pm$ & 1.005 $\pm$ \\ 
     & & & & & & 20.1 & 5.96e-3  &  2.97e-3  & 3.08e-3  \\
    & & & & No &  1.3 & 104.2 $\pm$  & 5.97e-2 $\pm$ & 1.704 $\pm$  &  1.003 $\pm$ \\ 
    & & & &  & & 21.9 & 6.47e-3 &3.14e-3 & 3.24e-3 \\ 
    & & & $\{ \beta_0\}$ & Yes & 1.5 & 102.8 $\pm$ & 5.93e-2 $\pm$  & 1.699 $\pm$ &  -  \\
    & & & &  & & 16.5  &5.1236e-3  & 2.71e-3 &  \\ 
    & & & & No & 1.9 &  114.5 $\pm$ &  6.25e-2 $\pm$ & 1.707$\pm$   &  - \\ 
    & & & &  & & 15.6 &4.96e-3 & 2.59e-3 &   \\ 
    && Inverted &  $\{ \beta_0, \beta_1\}$ & Yes &  0.0  & \multirow{2}{*}{$113.0^{+9.06 *}_{-0.72}$}  & 6.30e-02 $\pm$  &   1.700 $\pm$  & 9.99e-1 $\pm$  \\
        & & & &  & &  & 3.34e-03  & 3.07e-03  &   2.64e-03 \\ 
     && Normal &  $\{ \beta_0, \beta_1\}$ & Yes & 0.9 & 98.90 $\pm$  & 5.89e-2 $\pm$  & 1.701 $\pm$  & 1.00 $\pm$   \\
         & & & &   & &  21.3 & 6.18e-3 & 3.13e-3 & 3.09e-3  \\ 
     \cline{2-10}
     & CMB-S4  & Degenerate & $\{ \beta_0, \beta_1\}$ & Yes & 1.3 & 102.9 $\pm$ & 5.95e-2 $\pm$ & 1.699 $\pm$ & 1.001 $\pm$\\ 
         & & & &  & & 27.5  &  8.29e-3 &   3.31e-3 & 2.94e-3  \\ 
\botrule     
\end{tabularx}
    \caption{Comparison of reconstructed mean and error of cosmological and nuisance parameters as well as best-fit log-likelihoods with respect to the fiducial for different models, with {\it Euclid} mock data. The fiducial for all these cases is the same and is given in Table~\ref{tab:fiducials}, which is exactly recovered by the inverted model tabulated here. The $\Sigma m_ \nu$ posterior of this model (denoted by an asterisk), is truncated by the prior at its minimum mass and thus is narrower than its normal and degenerate hierarchy counterparts.}
    \label{tab:postparams_eu}
\end{table*}

\begin{table*}[]
\begin{tabularx}{\textwidth}{ c | c | c |  c | C | C | C | C | C }
    \toprule
     \multicolumn{2}{c|}{Data} & \multicolumn{3}{c|}{Model} & 
     \multicolumn{4}{c}{Mean and error} \\
     \hline 
      LSS & CMB & Hierarchy & Nuisance & GISDB  & $-2 \Delta\log \mathcal{L}$ & $\sum m_\nu$ [meV] & $\taur$ & $\beta_0 $    \\
     \hline 
  \multirow{9}{*}{DESI }  & \multirow{8}{*}{CMB-S4 + $\taur$}   & Degenerate  &$\{ \beta_0 \}$ & Yes &  0.9 & 107.6 $\pm$  & 5.99e-2 $\pm$  & 1.000 $\pm$  \\ 
     & & & &   & & 26.7 & 7.20e-3 & 1.70e-3   \\ 
    & & & & No & 1.1 & 112.0 $\pm$ &  6.07e-2 $\pm$ & 1.003 $\pm$  \\ 
    & & & &  & & 26.1 & 6.93e-3 &1.73e-3  \\ 
    && Inverted &  $\{ \beta_0\}$ & Yes &  0.0 &  \multirow{2}{*}{$107.2^{ +15.2 *}_{-0.42}$}   & 6.16e-2 $\pm$   &  1.001 $\pm$  \\
      & & & & &  &   & 3.84e-3 & 1.63e-3 \\
     && Normal &  $\{ \beta_0\}$ & Yes & 1.0  & 99.7 $\pm$  & 5.89e-2 $\pm$ & 1.000 $\pm$ \\
       & & & &  & &  28.6 & 6.52e-3 & 1.68e-3 \\
     \cline{2-9}
      & $\text{{\it Planck}}^\dag$ & Degenerate & $\{ \beta_0\}$ & Yes & -  & $\pm 27.44$  & $\pm 8.99$e-3 & $\pm 7.62$e-3 \\ 
\botrule     
\end{tabularx}
    \caption{Similar to Table~\ref{tab:postparams_eu}, with DESI used as LSS data. The fiducial for all these cases is the same and given in Table~\ref{tab:fiducials}. Note that as before, the $\Sigma m_ \nu$ posterior of the inverted model (denoted by an asterisk) is prior-informed and thus narrower than those of other hierarchies. In addition, the cases with {\it Planck} CMB data (denoted by a dagger) are reported as Fisher forecasts only, without an MCMC analysis.}
    \label{tab:postparams_desi}
\end{table*}

\subsection{Differentiation of Hierarchy}

We first consider whether we can differentiate the neutrino hierarchies, if they had the same $\sum m_\nu$. In Figure~\ref{fig:diff_hier} we show a corner plot comparing posteriors for the three hierarchies (the two physical ones plus the degenerate one), where the underlying fiducial cosmology is IH. 
Due to the physical lower bounds for the total mass of neutrinos in the IH, the posteriors for that case are notably one-sided, and as a result the mean value of most cosmological parameters for the other two hierarchies are shifted relative to the inverted one. 
This is because all cosmological parameters other than $\omega_b$ exhibit significant degeneracy with $\sum m_\nu$. However, as Table~\ref{tab:postparams_eu} shows, these near-future surveys show at most a $1\sigma$ preference for the fiducial choice of hierarchy, as far as best-fit likelihoods are concerned.
This agrees with the recent Bayesian analysis done in Ref.~\cite{Archidiacono:2020dvx}. 

\subsection{Effect of GISDB}

We then look more in detail into the runs with and without the GISDB. In this case the effects are twofold: a scale-dependent step that counteracts the scale-dependent suppression induced by neutrinos, and a redshift-dependent amplitude of the step that enhances the redshift-dependence induced by neutrinos at the smallest scales. Omission of the GISDB in the analysis analysis is then expected to underpredict $\sum m_\nu$ if the former effect is dominant, and overpredict if the latter effect is. However, the former effect is largely rendered insignificant due to cosmic variance -- the scale-dependence of the bias plateaus at scales smaller than  $\mathcal{O}(10^{-2} h\mathrm{Mpc}^{-1})$, which is the regime with strongest statistical power. Thus, it is the latter small-scale redshift-dependent effect that becomes relevant. 

\begin{figure*}
\begin{center}
\begin{subfigure}{0.40\linewidth}
\centering
    \includegraphics[width = \linewidth]{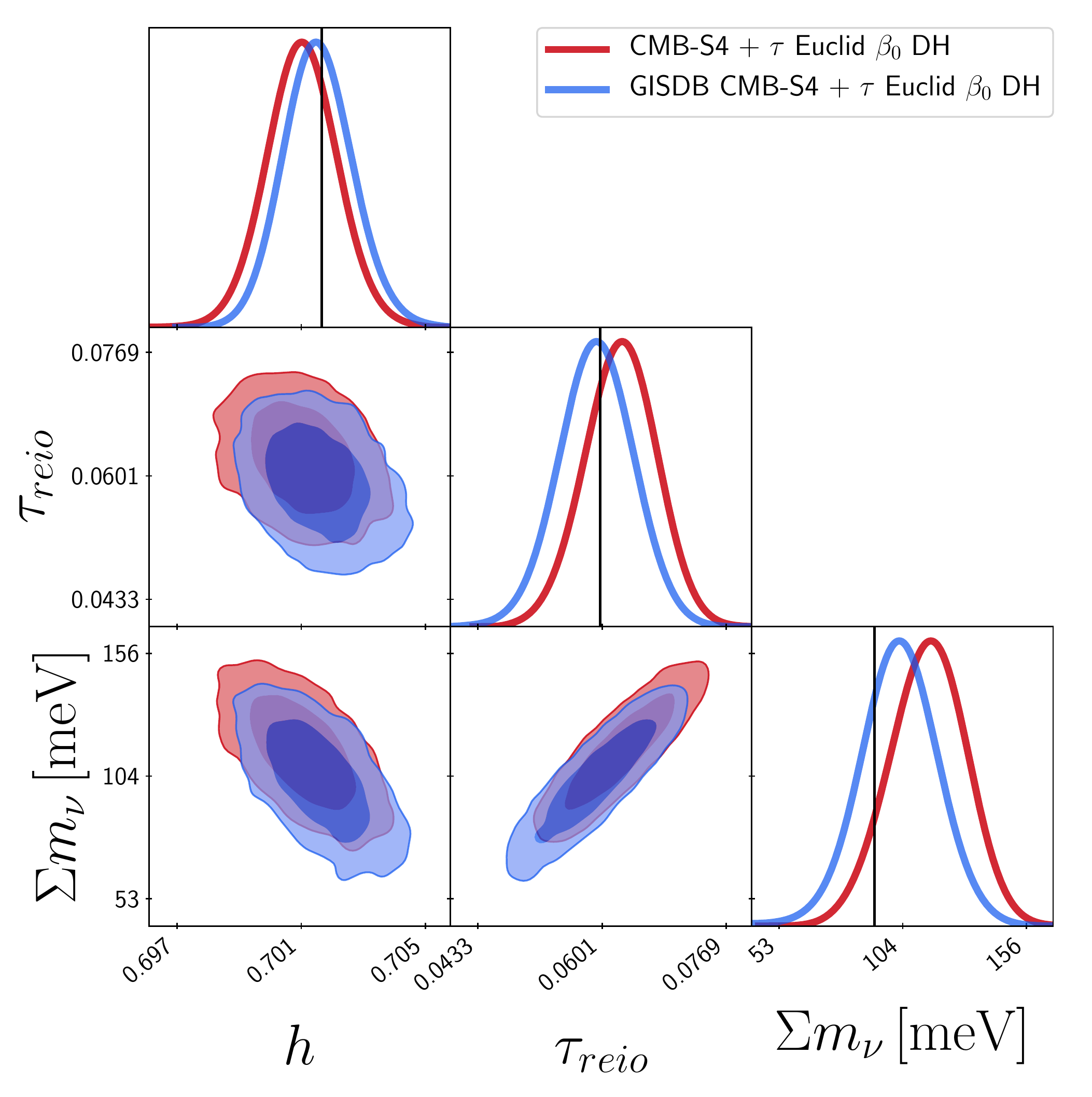}
    \caption{Posteriors for CMB-S4+{\it Euclid} with a prior on $\taur$ with a width of 0.01, fixing nuisance parameter $\beta_1$, assuming degenerate hierarchy (DH) with and without GISDB.
    Not marginalizing over $\beta_1$ produces a shift in key cosmological parameters due to the GISDB.
}
    \label{fig:diff_rf_3}
\end{subfigure}
\hspace{3mm}
\begin{subfigure}{0.40\linewidth}
\centering
    \includegraphics[width = \linewidth]{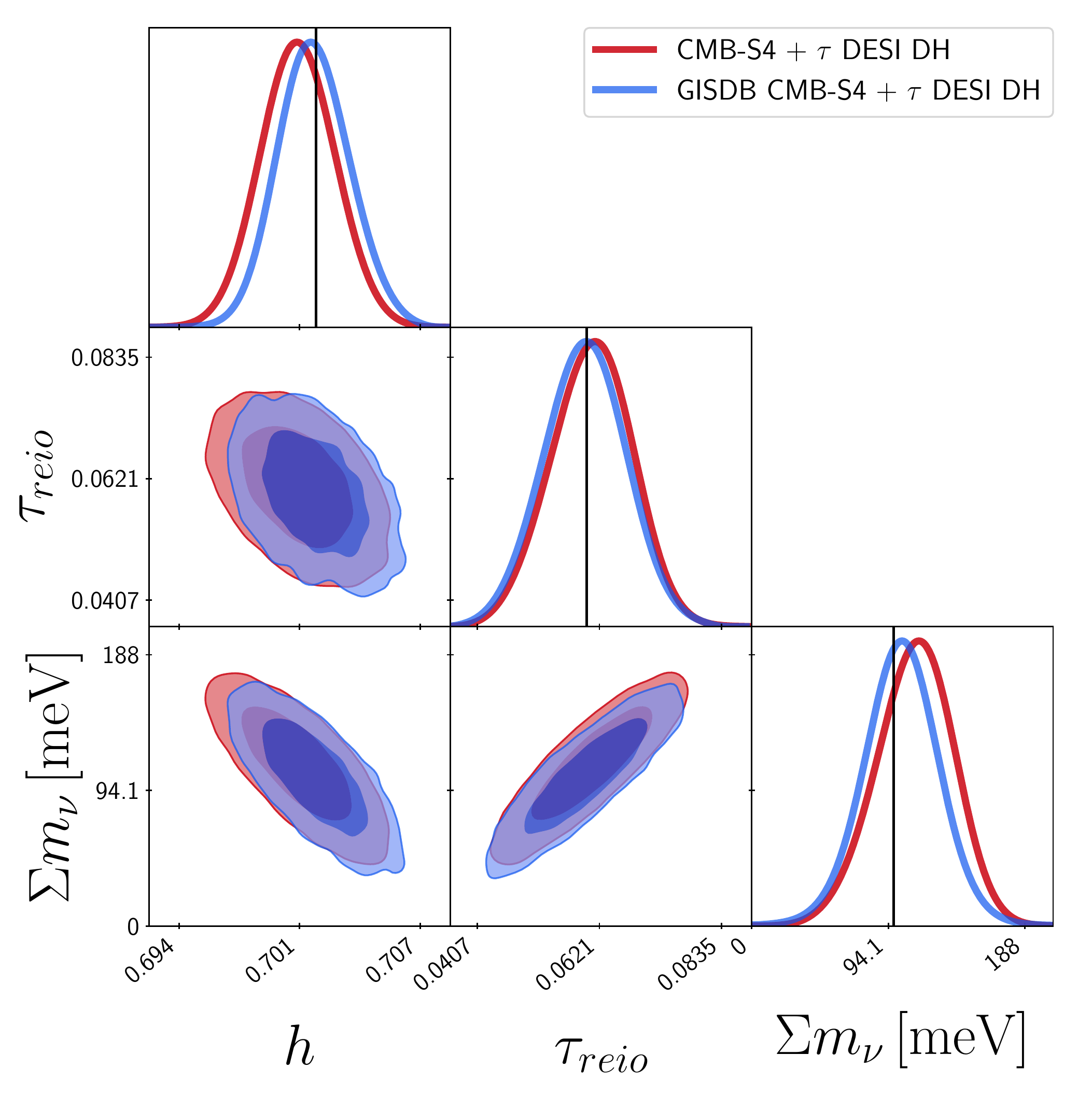}
    \caption{Posteriors for CMB-S4+DESI with a prior on $\taur$ with a width of 0.01, assuming degenerate hierarchy (DH) with and without GISDB.
    The DESI prescription of not marginalizing over the redshift dependence of the bias gives rise to small parameters shifts.
    }
    \label{fig:diff_rf_desi}
\end{subfigure}
\begin{subfigure}{0.40\linewidth}
\centering
    \includegraphics[width = \linewidth]{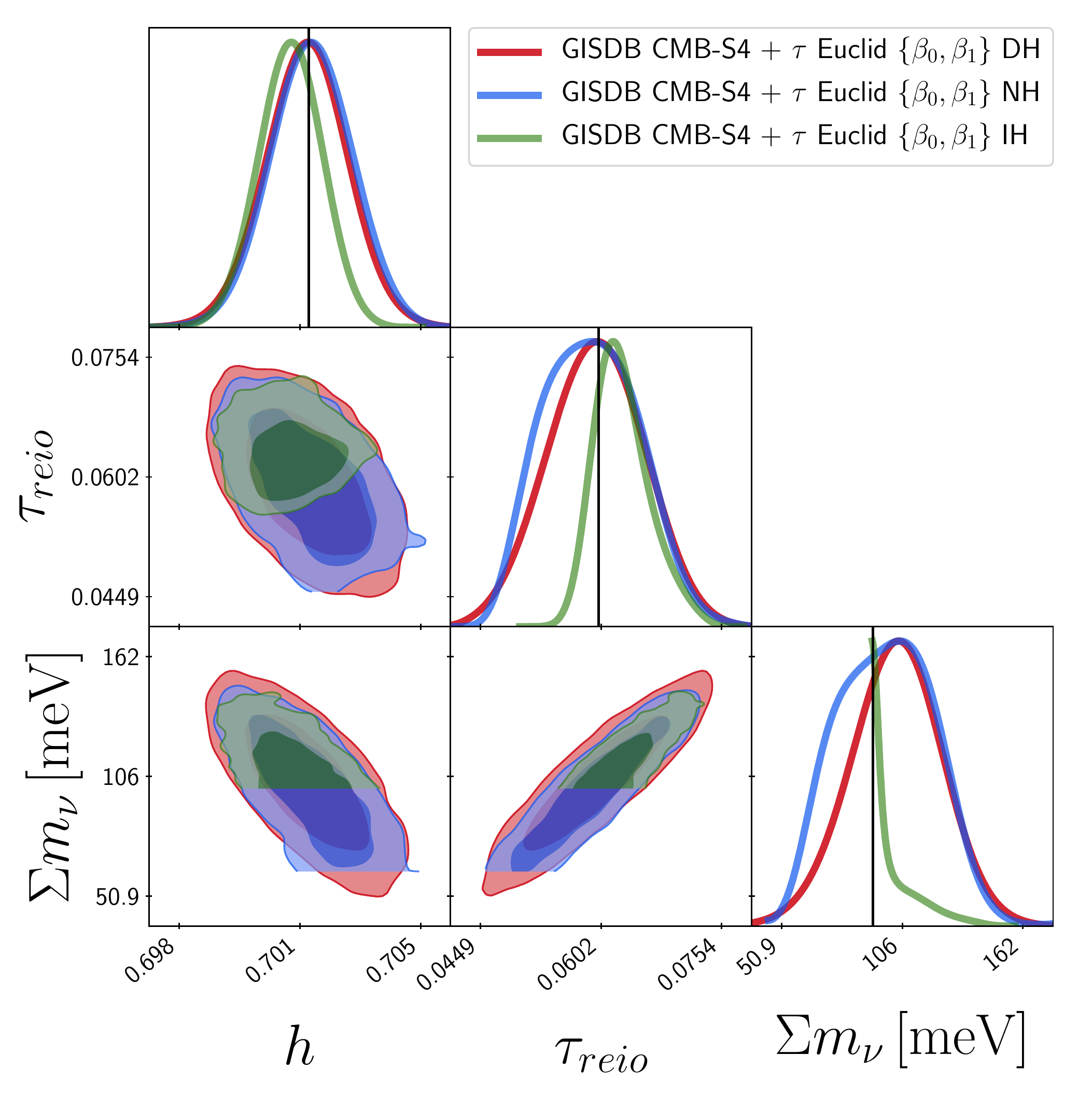}
    \caption{Posteriors for CMB-S4+{\it Euclid} with a prior on $\taur$ with a width of 0.01, assuming Degenerate (DH), Inverted (IH) and Normal (NH) hierarchies.
 }
    \label{fig:diff_hier}
\end{subfigure}
\hspace{3mm}
\begin{subfigure}{0.40\linewidth}
\centering
    \includegraphics[width = \linewidth]{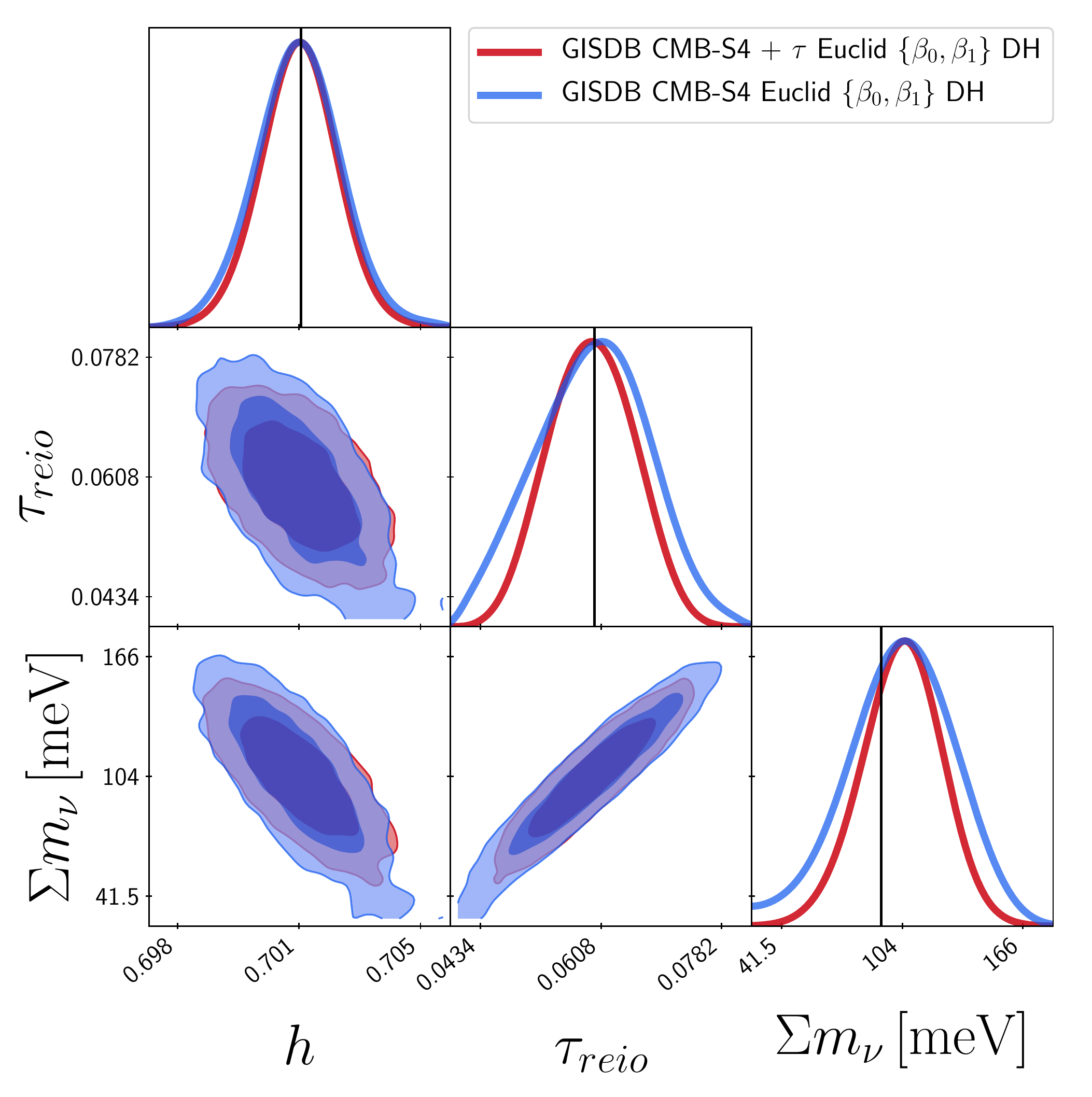}
    \caption{Posteriors for CMB-S4+{\it Euclid}, assuming Degenerate (DH), with and without a prior on $\taur$.
    }
    \label{fig:diff_taupri}
\end{subfigure}
\end{center}
\caption{MCMC Posteriors for $h$, $\tau_{\rm reio}$, and $\sum m_\nu$ for CMB-S4 and various LSS experiments. For each, the fiducial cosmology has an Inverted hierarchy in the minimum-mass configuration, with total neutrino mass 98 meV.
}
\label{fig:diff_rf_tots}
\end{figure*}

\begin{figure*}

    \includegraphics[width = \linewidth]{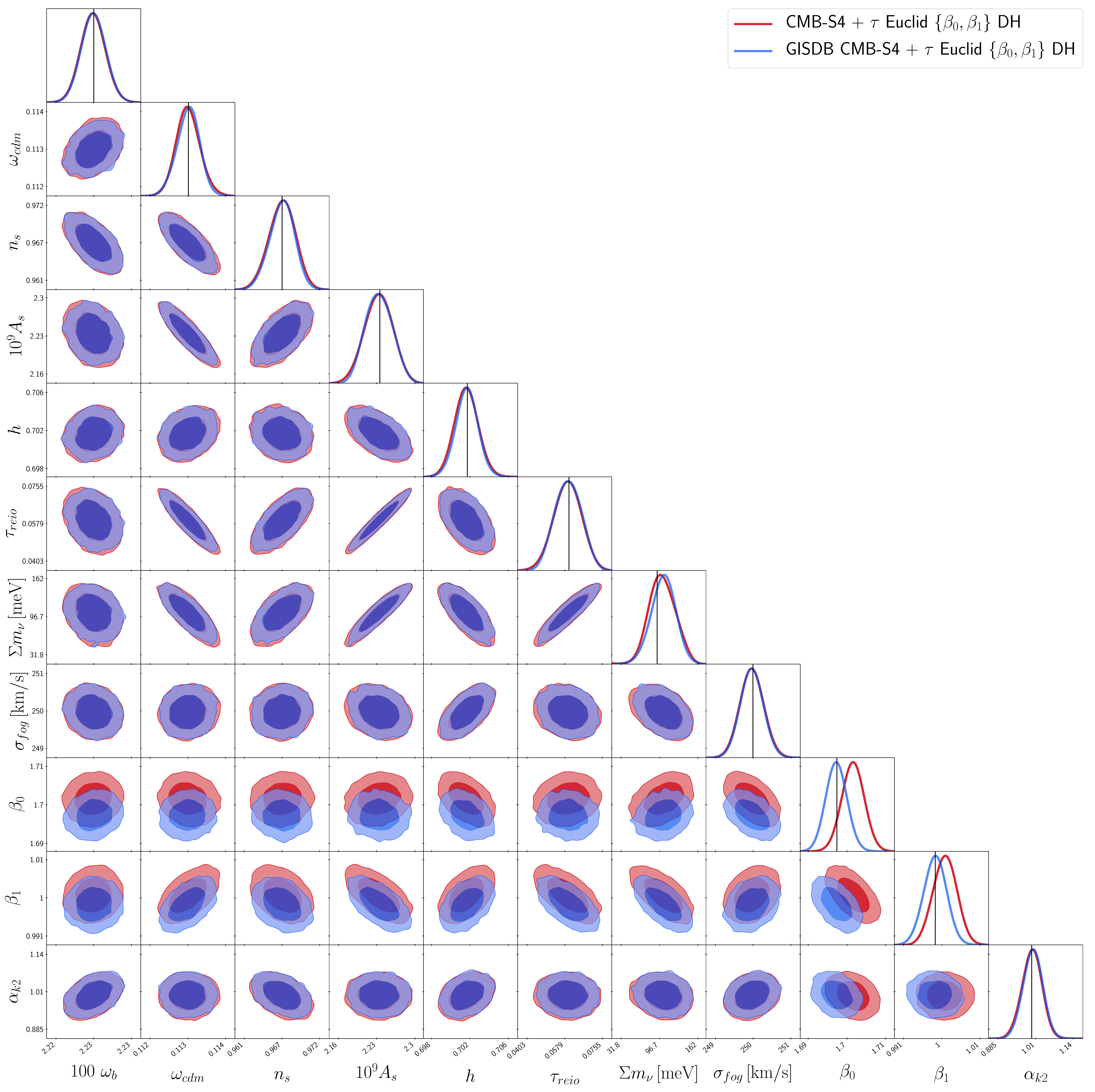}
    \caption{Posteriors for CMB-S4+{\it Euclid} with a prior on $\taur$ with a width of 0.01, assuming degenerate hierarchy (DH) with and without GISDB, and assuming the IH with total neutrino mass 98 meV as fiducial. 
    }
    \label{fig:diff_rf_2}
\end{figure*}

The most evident effect of neglecting this growth-induced step is a misreconstruction of the normalization bias, resulting in a shift towards larger $\bar b_L(z)$, and in turn the incorrect values of $\beta_0$ and its redshift dependence $\beta_1$, as seen in Figure~\ref{fig:diff_rf_2}.  
If the nuisance parameter $\beta_1$ was not marginalized over, this would further result in a significant shift of reconstructed cosmological parameters such as $A_s,\; h,$ and $\taur$, notably overestimating the total neutrino mass by $\gtrsim 1\sigma$, as expected. 
This effect is shown in Figure~\ref{fig:diff_rf_3}, where we show the ellipses with and without the GISDB, although as opposed to Figure~\ref{fig:diff_rf_2} we did not marginalize over $\beta_1$. 
As the DESI bias prescription does not include a degree of freedom to vary the redshift dependence, neglecting the GISDB induces a shift in the aforementioned cosmological parameters,
albeit at the sub-$\sigma$ level, due to the smaller signal-to-noise.
We illustrate this point in Figure~\ref{fig:diff_rf_desi}. 
We note that the same shifts can be recovered using a simpler Fisher-matrix formalism, as we show in our companion paper~\cite{DePorzio:2020wcz}.

In general, we expect the DESI and {\it Euclid} prescriptions for analysis of real data, when collected, to be more sophisticated than those presented in the Science Books. Nonetheless, this is additional reason for the analyses of upcoming e.g.~DESI data to marginalize over the redshift dependence of the bias, or parametrize the bias at each redshift bin independently, to avoid cosmological-parameter shifts due to the neutrino GISDB.
Note, however, that even when marginalizing over $\beta_1$ there is a leftover shift on the scale-independent bias $\beta_0$, as seen in Figure~\ref{fig:diff_rf_2}.
The value of this parameter affects other observations, such as galaxy high-order functions and cross correlations with other data sets, so if one requires an unbiased estimate of $\beta_0$ the full GISDB ought to be accounted for.

\section{Conclusions}
\label{sec:conclusions}

In this work we presented forecasts on the ability of current and upcoming CMB and LSS experiments to measure neutrino masses, both in total and individually.
We included all known linear effects induced by neutrinos in the treatment of galaxy survey data,  specifically assessing the impact of the scale-dependent bias induced by the effect of neutrinos in the growth function. We also investigated the effect and detectability of realistic neutrino hierarchies in the analysis of these survey data.

Starting with the different neutrino hierarchies, we have shown that for the data considered, the total neutrino mass is determined up to an uncertainty of $\sigma \sum m_\nu \sim 20$ meV at a fiducial of $ \sum m_\nu =98$ meV, the minimum-mass scenario of the inverted hierarchy. While this is a $5\sigma$ detection away from 0, the minimum-mass scenario of the normal hierarchy ($ \sum m_\nu =60$ meV) is excluded only at the $2\sigma$ level. Furthermore, for a fixed total neutrino mass of $98$ meV, a different choice of hierarchy constitutes a difference of $\sim 0.3\%$ in the power spectrum amplitude, and we show that this is expected to result in a $1\sigma$ shift in inferred cosmological parameters. As such, more advanced surveys are necessary to definitively distinguish between the two hierarchies, particularly in the case that the total neutrino mass is the same.

As for the growth-induced scale-dependent bias (GISDB), we find that, while cosmic variance limitations render the data insensitive to the scale-dependent shape of the halo bias for allowable neutrino masses, upcoming surveys are expected to be highly sensitive to the redshift dependence of the GISDB at small scales. 
In joint analyses of CMB-S4 data with large-scale surveys such as DESI or {\it Euclid}, not including this GISDB step can result in a $\mathcal{O}(1\sigma)$ over-prediction of total neutrino masses, as well as similarly shifted reconstructions for degenerate parameters such as $h$ and $\taur$. If one marginalizes over the redshift dependence of the bias, these shifts can be removed, although the resulting analysis will retain a $\mathcal{O}(1\sigma)$ shift in the magnitude and redshift dependence of the Eulerian bias of the relevant tracers. 
Thus it is imperative to include these effects in order to accurately recover galaxy and cosmology parameters simultaneously, or otherwise marginalize over bias redshift dependence if the latter is prioritized.

\section*{Acknowledgments}

We are grateful to Francis-Yan Cyr-Racine, Ana Diaz Rivero, and Sunny Vagnozzi for helpful comments on a previous version of this manuscript.
ND was supported by a National Physical Science Consortium fellowship.
CD and JBM were partially supported by NSF grant AST-1813694.

\bibliography{neutrino_forecasts}

\end{document}